\documentclass[final,1p,times]{elsarticle} 
\usepackage{graphicx} 
\usepackage{amssymb} 
\usepackage{amsthm} 
\usepackage{lineno} 

\journal{Nuclear Physics A} 
\begin{document} 

\begin{frontmatter} 


\title{Direct $\gamma$ and $\gamma$-jet measurement capability of ATLAS \\ 
for Pb+Pb collisions}

\author{Mark D. Baker$^a$ for the ATLAS Collaboration} 

\address[a]{Physics Department, Brookhaven National Laboratory,
Upton, NY 11973-5000}
\begin{abstract} 
%
The ATLAS detector at the LHC is capable of efficiently separating
photons and neutral hadrons based on their shower shapes over a wide
range in $\eta$, $\phi$, and $E_{\rm T}$, either in addition to or
instead of isolation cuts. This provides ATLAS with a unique strength
for direct photon and $\gamma$-jet physics as well as access to the
unique capability to measure non-isolated photons from fragmentation
or from the medium. We present a first look at the ATLAS direct photon
measurement capabilities in Pb+Pb and, for reference, p+p collisions
at $\sqrt{s_{_{NN}}}=5.5$~TeV over the region \mbox{$|\eta|<2.4$}.

\end{abstract} 

\end{frontmatter} 



\section{Introduction}

Direct photons produced in Pb+Pb collisions can be divided into prompt
photons produced in hard processes in the initial collision, and
non-prompt photons produced by jet fragmentation, in-medium gluon
conversion and medium-induced bremsstrahlung. Prompt processes such as
\mbox{$q+g \rightarrow q+\gamma$} and \mbox{$q+\bar{q} \rightarrow
g+\gamma$} lead to final states with a high $p_{\rm T}$ parton (gluon
or quark) balanced by a prompt photon with roughly comparable $p_{\rm
T}$~\cite{incnll}. They thus provide {\em a calibrated parton} inside
of the medium, allowing a direct, quantitative measurement of the
energy loss of partons in the medium and of the medium response.

ATLAS has a unique capability to study such processes because of the
large-acceptance calorimeter with longitudinal and fine-transverse
segmentation~\cite{ATLAScal}. In particular the first main layer of
the calorimeter is read out in narrow transverse strips. This
segmentation allows us to purify our sample of $\gamma$-jet events by
rejecting jet-jet background. It further allows us to identify photons
which are near or even inside of a jet, where isolation cuts cannot be
used. This provides access to non-prompt photons from jet
fragmentation, from in-medium gluon conversion and from the
medium-induced bremsstrahlung.

\section{Technique}

\begin{figure}[ht]   
\begin{center}
\includegraphics[width=0.31\textwidth]{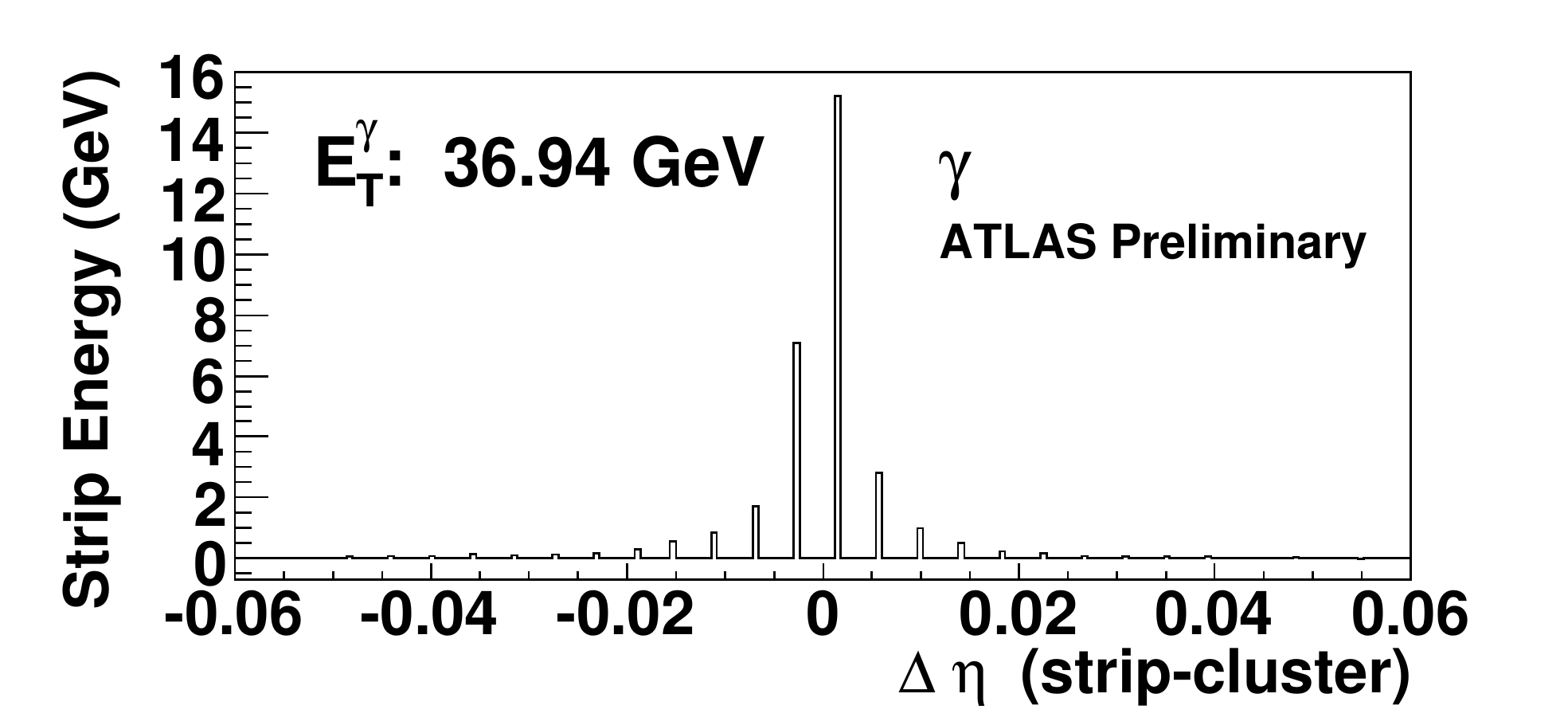}
\includegraphics[width=0.31\textwidth]{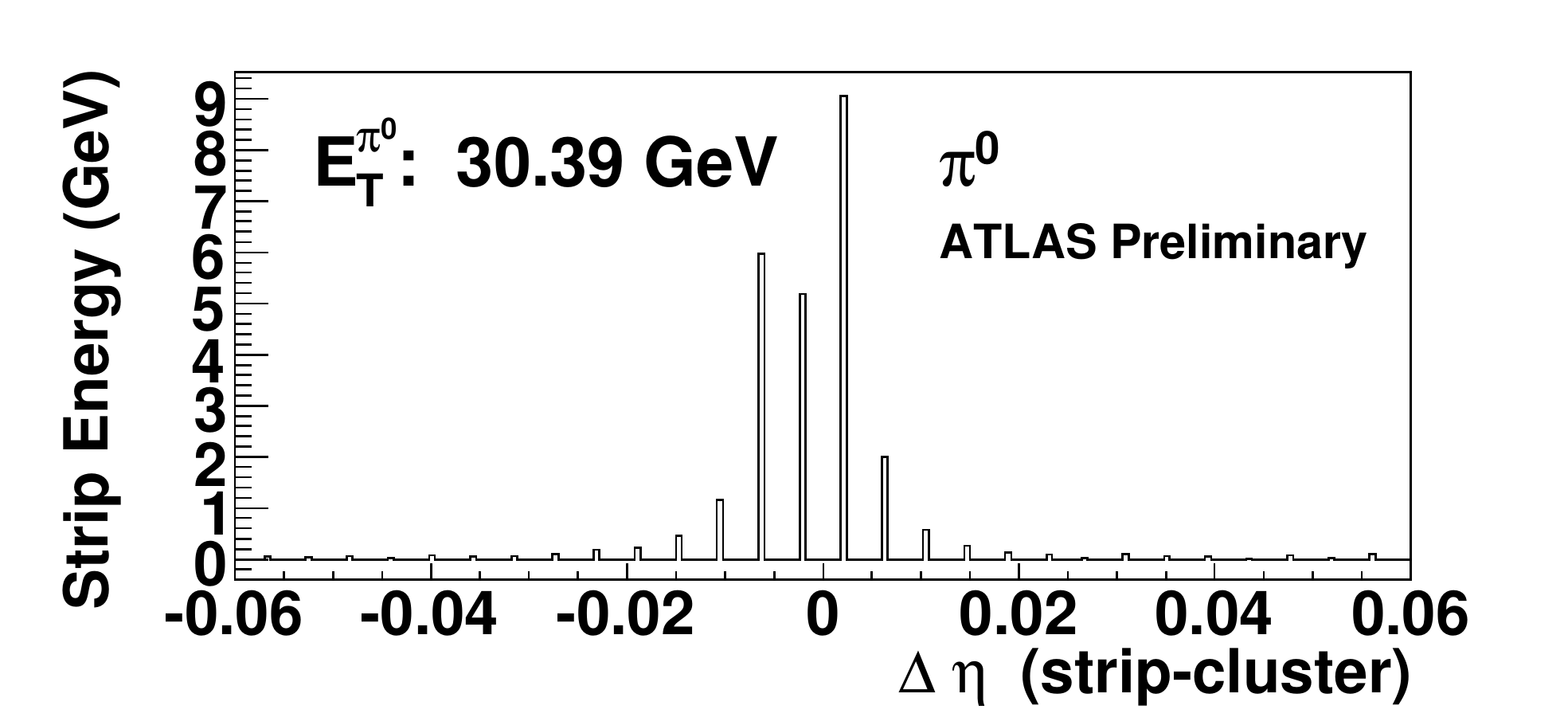}
\includegraphics[width=0.31\textwidth]{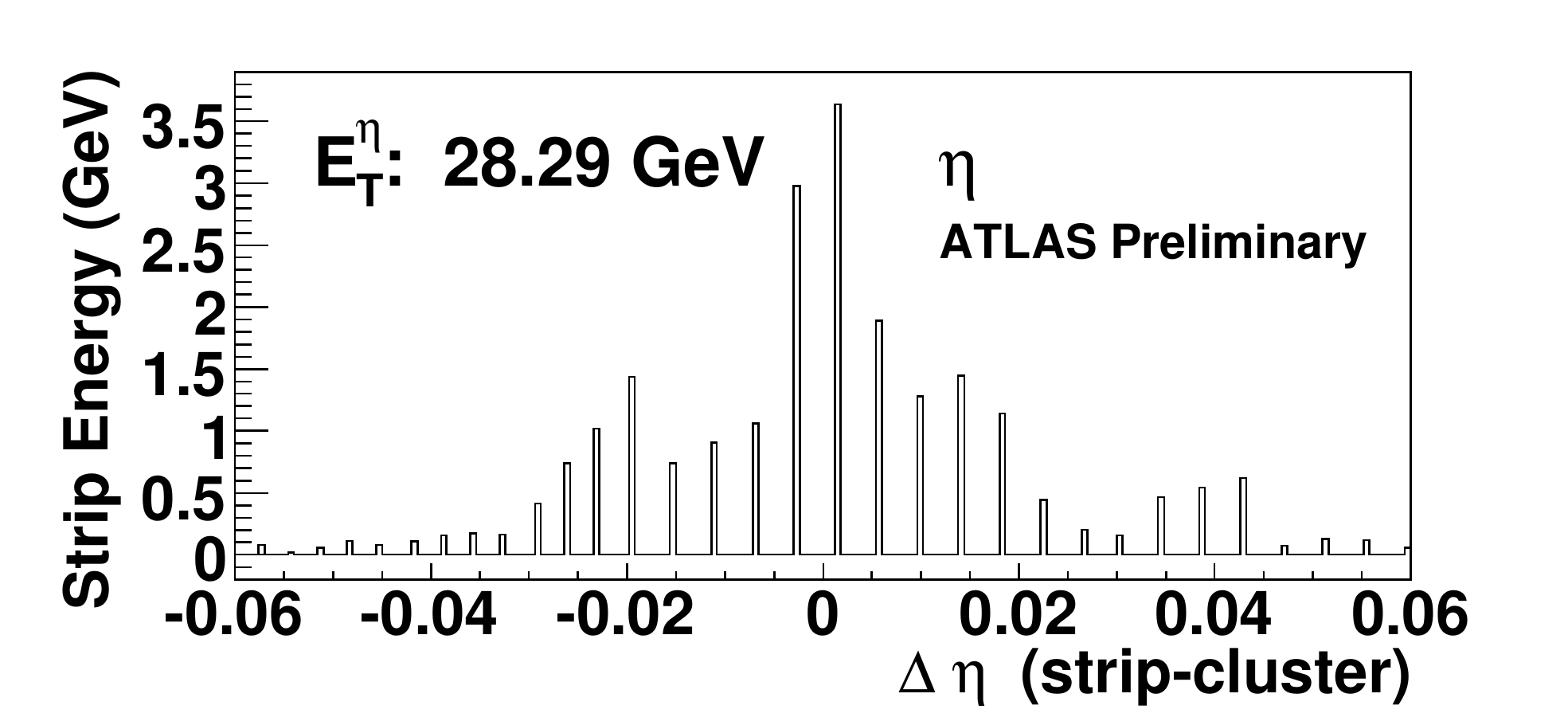}
\includegraphics[width=0.31\textwidth]{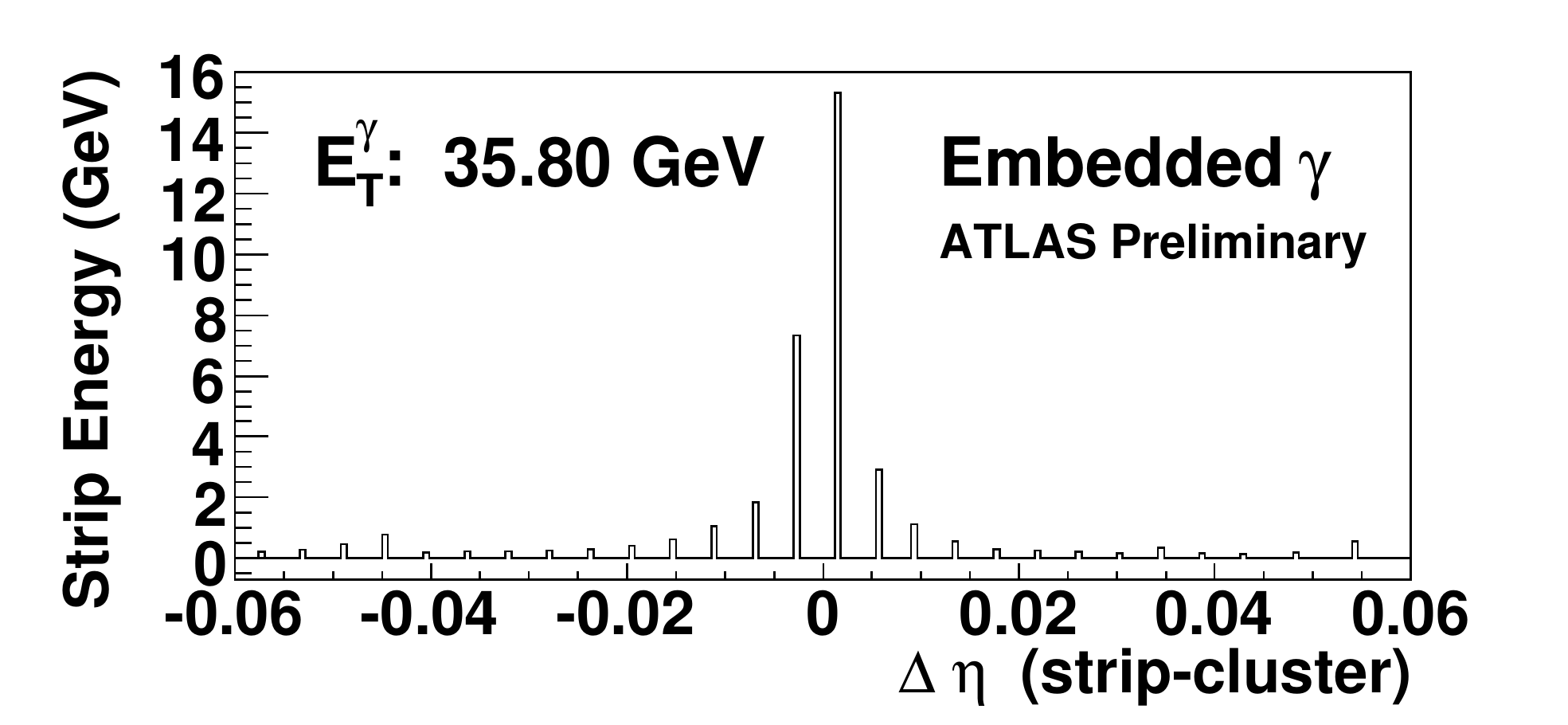}
\includegraphics[width=0.31\textwidth]{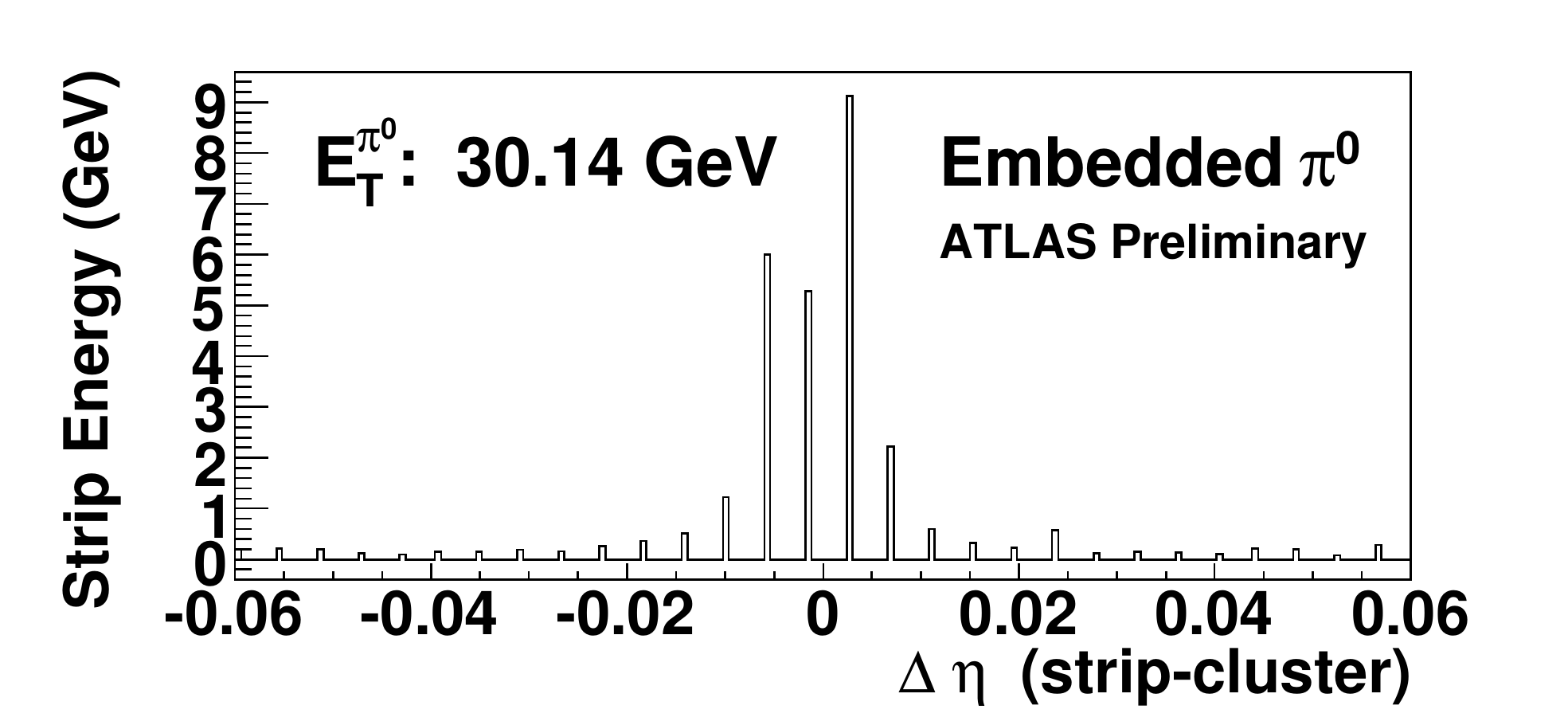}
\includegraphics[width=0.31\textwidth]{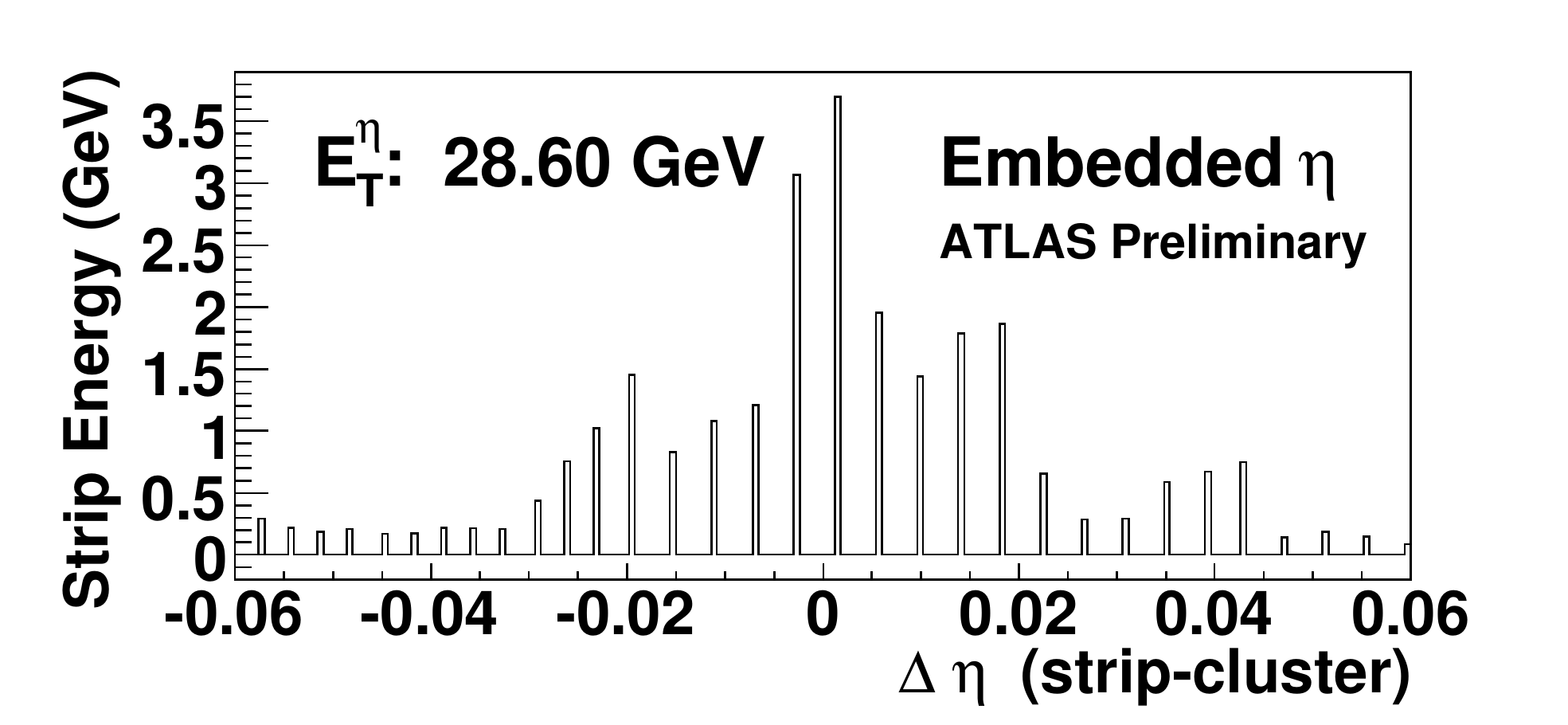}
\caption[]{\label{fig:strip6} The energy deposition in the
strip layers around the direction of (upper left) a single photon,
(upper middle) a single $\pi^0$ and (upper right) a single $\eta$ as
well as for (lower panels) the identical particles embedded in a
central ($b=2$~fm, $dN_{ch}/d\eta=2650$) Pb+Pb event. Reconstructed $E_{\rm T}$
values are indicated.}
\end{center}
\end{figure}

The design of the ATLAS electromagnetic calorimeter is optimal for
direct photon identification. The first layer of the electromagnetic
calorimeter, which covers the full azimuth and $|\eta|<2.4$, has very
fine segmentation along the $\eta$ direction (ranging from 0.003 to
0.006 units). This layer provides detailed information on the shower
shape, which allows a direct separation of $\gamma$'s, $\pi^0$'s, and
$\eta$'s on a particle-by-particle level. Deposited strip energy
distributions as a function of eta relative to the cluster centroid
for a typical single $\gamma$, single $\pi^0$, and single $\eta$ meson
are shown in the upper panels of Fig.~\ref{fig:strip6}.
Characteristically different shower profiles are seen. The energy of a
single photon is concentrated across a few strips, with a single
maximum in the center, while the showers for $\pi^0\rightarrow
\gamma\gamma$ and $\eta\rightarrow \gamma\gamma$ are distributed
across more strips, often with two or more peaks.  The broad shower
profile for $\pi^0$ and $\eta$ reflects the overlap of showers for two
or more decay photons.  Even when the two peaks are not resolved, the
multi-photon showers are measurably broader on a statistical basis.
The lower panels of Fig.~\ref{fig:strip6} show the strip layer energy
distributions surrounding the direction of single particles embedded
in central Pb+Pb events. The $\gamma$, $\pi^0$ and $\eta$ in these
panels are the same ones used in the upper panels. Despite the large
background of low-energy particles produced in Pb+Pb
events~($dN_{ch}/d\eta=2650$ in this case), the shower shape for the
embedded particle is almost unchanged by the background.  Thus the
strip layer allows the rejection of $\pi^0$ and $\eta$ clusters over a
very broad energy range, and the performance for the background
rejection and identification efficiency should not depend strongly on
the event centrality.

\section{Results}

To distinguish direct photons from neutral hadrons, cuts have been
developed based on the shower shape in the strip layer. These cuts
reject those showers that are anomalously wide or exhibit a double
peak around the maximum. In general, better rejection can be achieved
using a tighter cut, but at the expense of reduced efficiency.  The
performance has been quantified via photon efficiency
($\epsilon_{\gamma}$) and relative rejection ($R_{\rm rel} \equiv
\epsilon_{\gamma}/\epsilon_{\rm hadron}$).  The relative rejection
basically reflects the gain in the signal (direct photon yield)
relative to background (neutral hadron yield).

\begin{figure}[ht]
\begin{center}
\includegraphics[width=0.75\linewidth]{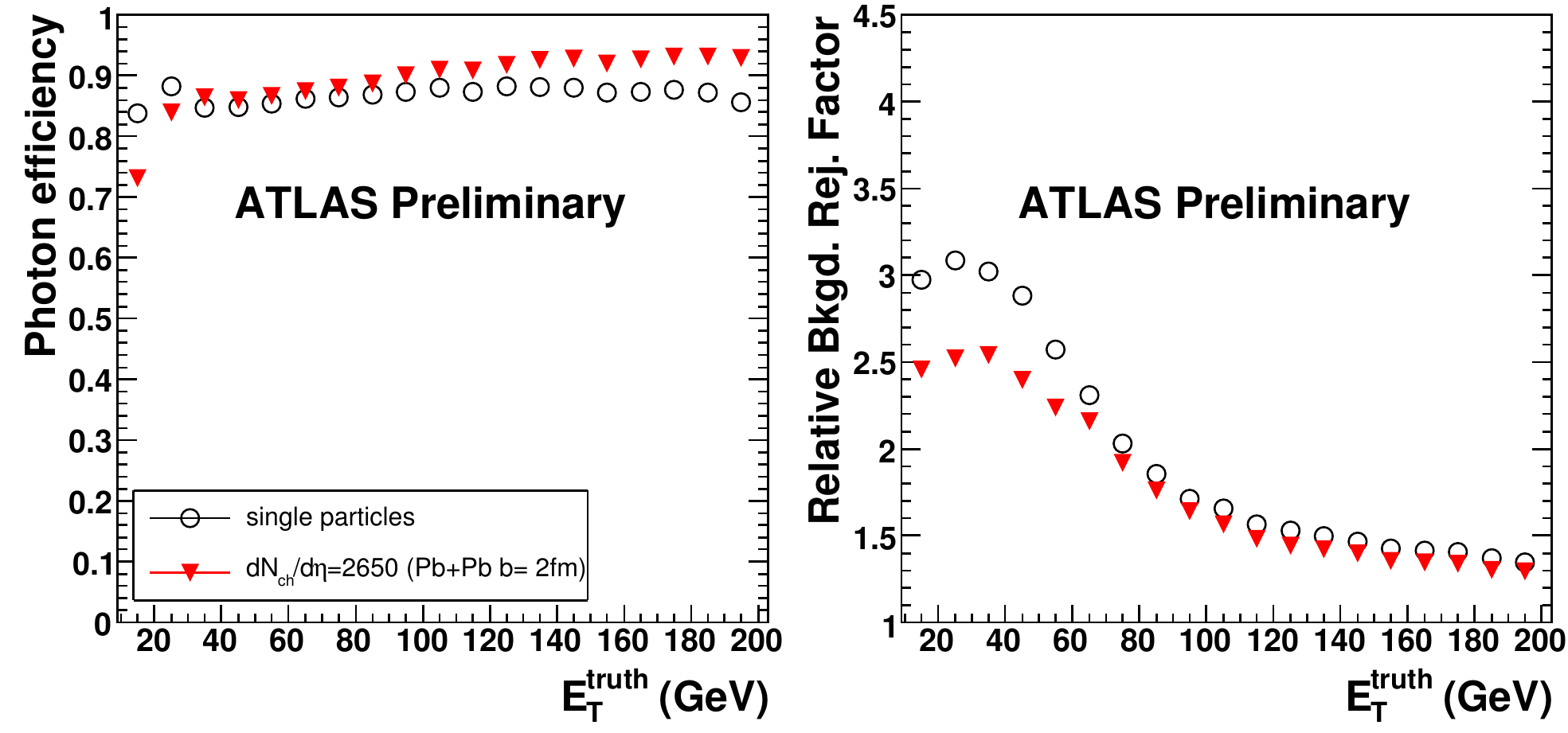}
\includegraphics[width=0.75\linewidth]{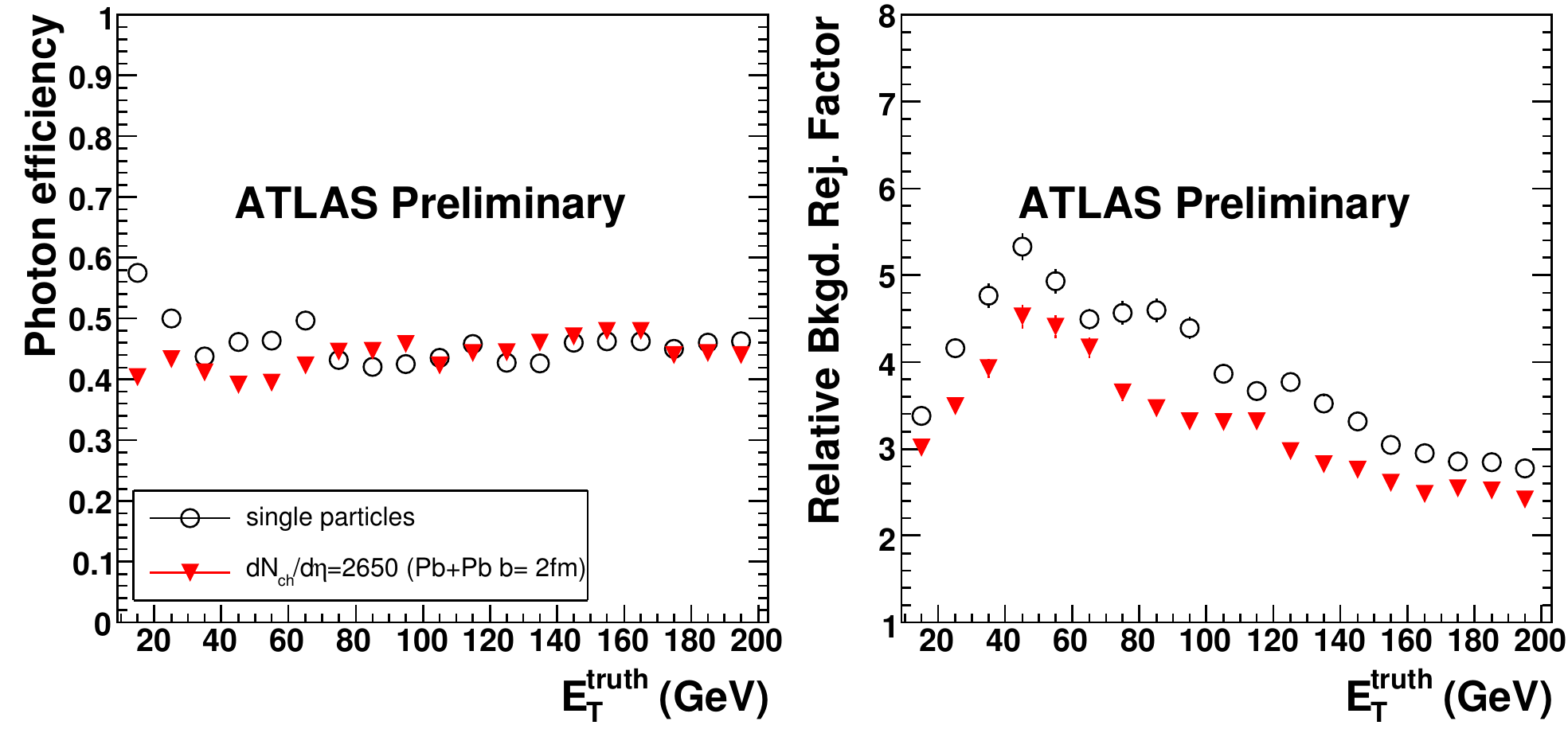}
\caption[]  {\label{fig:bothcuts} (upper panels) Photon identification 
efficiency and relative rejection factor (averaged over $|\eta|<2.4$)
for neutral hadrons for the loose cut set for single particles (open
circles) and central ($b=2$~fm, $dN_{ch}/d\eta=2650$) Pb+Pb collisions
(filled triangles). (lower panels) As above but for the tight cut set.
Note the change in scale between the upper and lower right-hand
panels.}
\end{center}
\end{figure}

In this analysis, two sets of cuts have been developed, a ``loose''
cut set and a ``tight'' cut set.  The performance for these two sets is
summarized in Fig.~\ref{fig:bothcuts}.  The loose cuts (upper panels)
yield a factor of 1.3--3 relative rejection with a photon efficiency
of about 90\%; the tight cuts (lower panels) yield a factor of 2.5--5
relative rejection with an efficiency of about 50\%.

\begin{figure}[h!t]
\begin{center}
\includegraphics[width=1.0\linewidth]{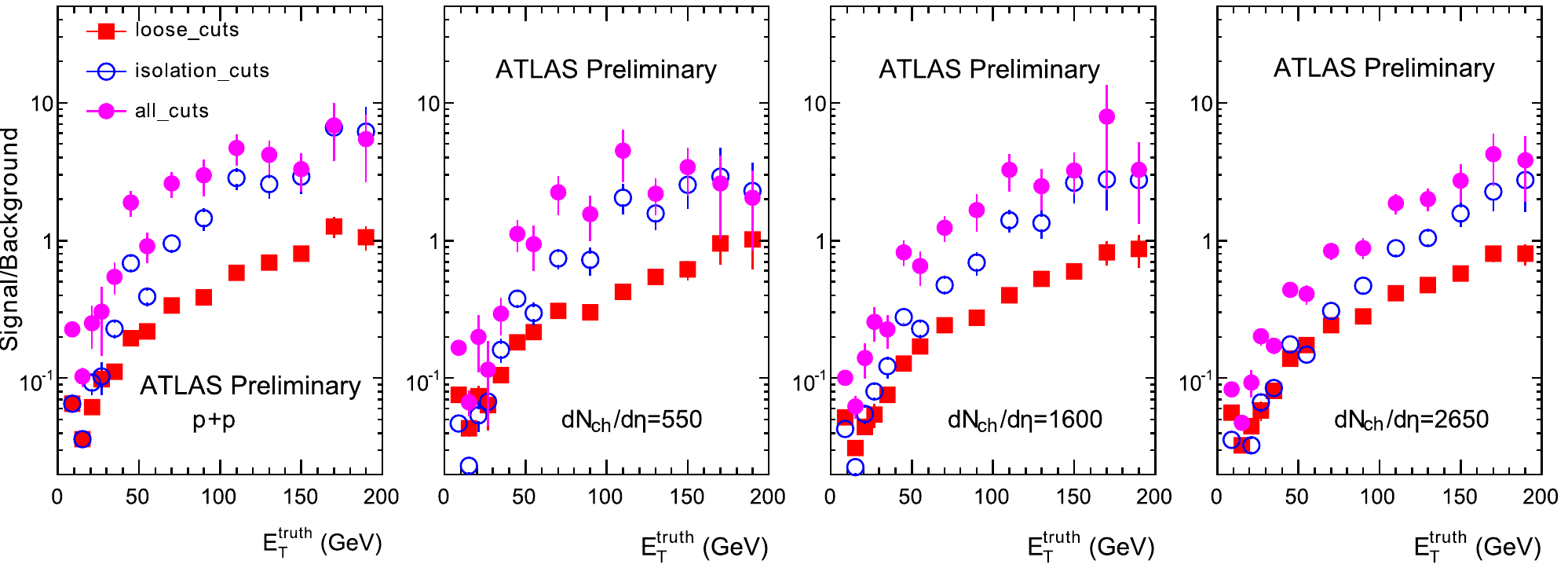}
\caption[]
{\label{fig:rejsncent} The ratio of direct photons over background
neutral hadrons passing the loose shower shape cuts only (solid
squares), isolation cuts only (open circles) and combined cuts (solid
circles) for different occupancies under the assumption that there is
no hadron suppression for any centrality.}
\end{center}
\end{figure}

In addition to the photon identification cuts, isolation cuts have
been developed which, on their own, provide relative rejection factors
of 7--10 for \mbox{$E_{\rm T}>50$~GeV}. These isolation cuts cannot be
used to study non-isolated photons, but in the case of $\gamma$-jet,
they can be combined with the photon identification cuts to
significantly reduce the background from jet-jet
events. Figure~\ref{fig:rejsncent} shows the signal-to-background
ratio after applying the loose shower shape cuts, the isolation cuts,
and the combined cuts. The signal-to-background ratio is the best in
p+p collisions, which is about factor of 4--5 larger than that for
most central Pb+Pb events. However, by taking into account the benefit
one gains from the likely hadron suppression~($R_{AA}=0.2$), we expect
to achieve a similar level of performance that is approximately
independent of the event centrality.
 
\begin{figure}[h!t]
\begin{center}
\includegraphics[width=0.49\linewidth]{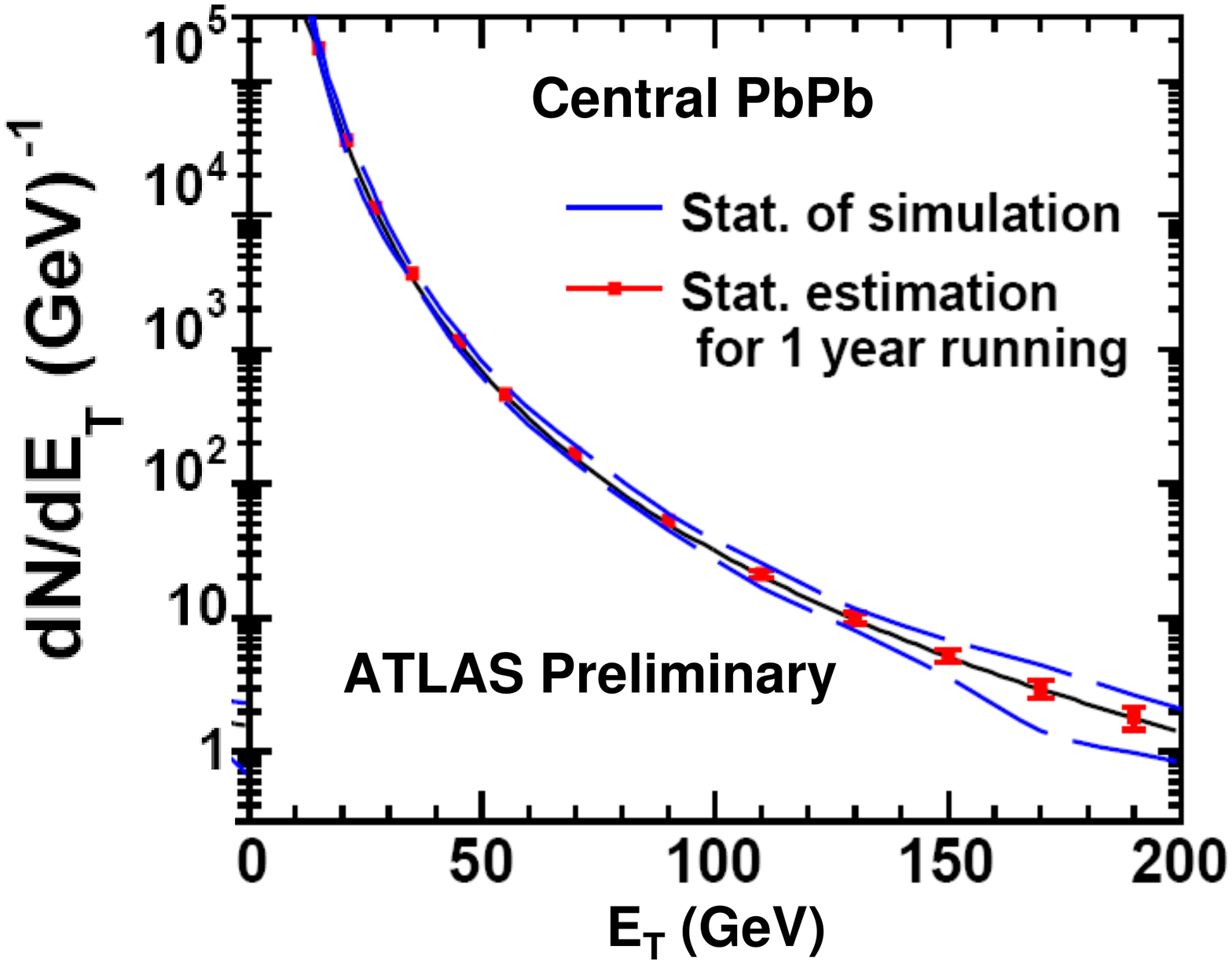}
\includegraphics[width=0.50\linewidth]
		{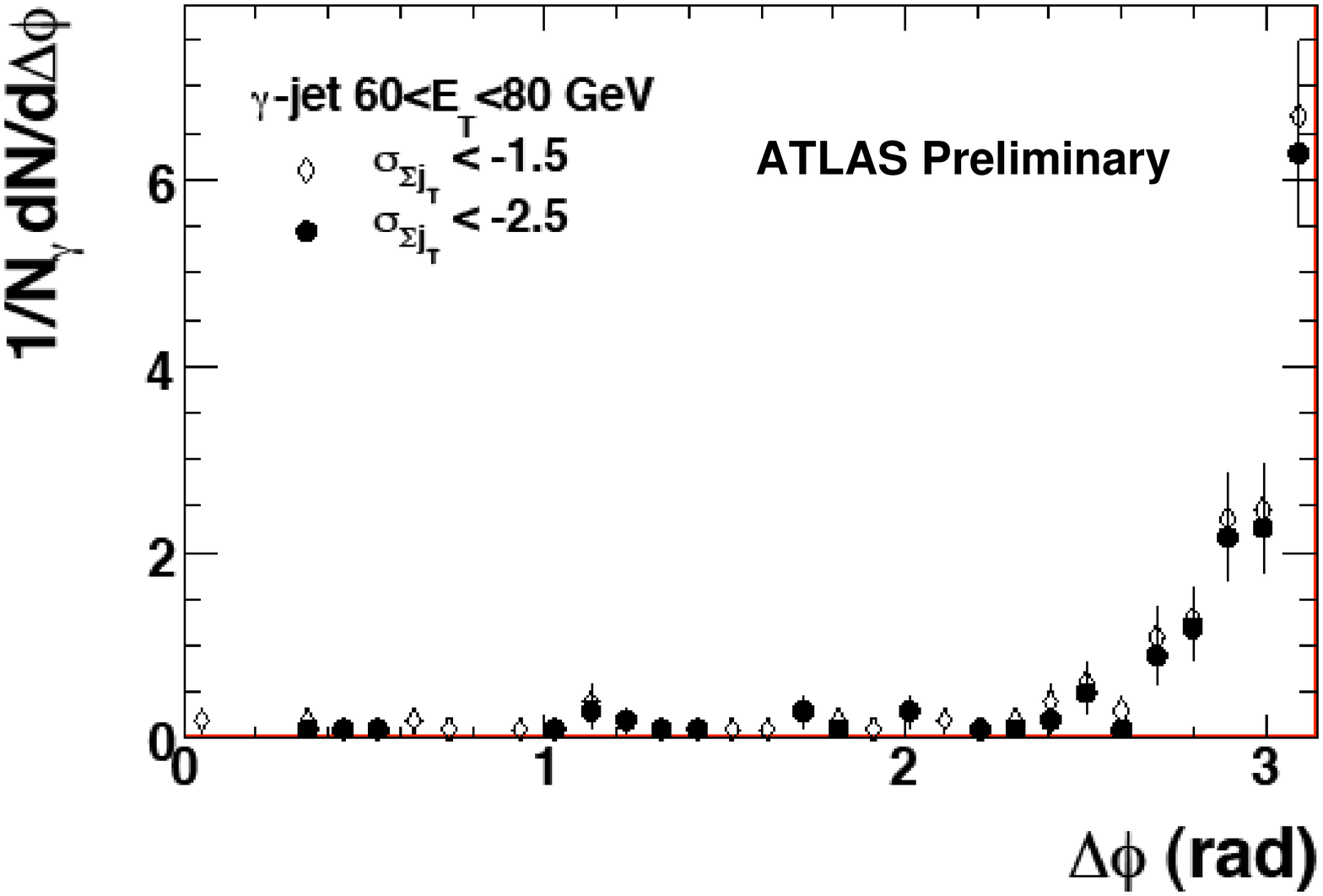}
\caption[]
{\label{fig:phospect} (left panel) A simulated photon spectrum is
shown along with expected statistical error bars after background
subtraction for a central 10\% Pb+Pb sample with $dN_{ch}/d\eta=2650$
from a nominal Pb+Pb run. (right panel) Correlations in $\Delta\phi$
for $\gamma$-jet pairs embedded in central Pb+Pb events, where both the
photon and jet have an $E_{\rm T}$ of 60--80~GeV. Filled circles refer to
jets passing a tighter jet quality cut than those represented by the
open circles.}
\end{center} 
\end{figure}

The left-hand panel of Fig.~\ref{fig:phospect} shows the performance
for reconstructing the direct photon spectrum for a central Pb+Pb data
sample, indicating that the spectrum can be measured out to at least
200~GeV at the expected luminosity per LHC Pb+Pb year (0.5 nb$^{-1}
\times $50\%). The right-hand panel shows the $\gamma$-jet correlation
for 60--80~GeV photons and jets in central Pb+Pb collisions (without
jet quenching or modification). For more details on the jet
reconstruction, see Ref.~\cite{jets}.
 
\section{Conclusions}

This writeup has presented the ATLAS performance for direct photon
identification. The first layer of the ATLAS electromagnetic
calorimeter provides an unbiased relative rejection factor of either
1.3--3 (loose shower shape cuts) or 2.5--5 (tight shower shape cuts)
for neutral hadrons.  The loose $\gamma$ identification cuts can be
combined with isolation cuts, resulting in a total relative rejection
of about 20, even in central Pb+Pb collisions, providing a relatively
pure sample of calibrated partons interacting with the medium.  The
expected luminosity per LHC Pb+Pb year (0.5 nb$^{-1} \times $50\%)
will provide 200k photons above 30~GeV, and 10k above 70~GeV per LHC
year.

The tight shower shape cuts alone provide sufficient rejection against
hadron decays within jets to allow the study of fragmentation photons,
in-medium gluon conversion and medium-induced bremsstrahlung.  This
capability combined with a large acceptance is unique to ATLAS.





\begin{thebibliography}{00} 
   
\bibitem{incnll} P.Aurenche, R.Baier, M.Fontannaz, and D.Schiff,
``Prompt Photon Production at Large $p_{\rm T}$: Scheme Invariant QCD
Predictions and Comparison with Experiment'', Nucl.\ Phys.\ {\bf B297},
(1988) 661.
\bibitem{ATLAScal} ATLAS Collaboration, G. Aad {\it et al.}, ``The
ATLAS Experiment at the CERN Large Hadron Collider'', 
Journal of Instrumentation 3 (2008) S08003.
\bibitem{jets} N.Grau {\it et al.}, ``Identification and Rejection of Fake
Reconstructed Jets From a Fluctuationg Heavy Ion Background in ATLAS'', 
ATL-PHYS-PROC-2008-046.
\end{thebibliography}
\end{document}